\documentclass{aastex}
\usepackage{ifthen}

\def \version {_working}

\ifthenelse{\equal{\version}{_apj}}
{

\def \figwidth {0.6 \linewidth}
}
{
\usepackage{emulateapj5}
\usepackage{apjfonts}
\usepackage{hyperref}

\def \figwidth {\linewidth}
}
\usepackage{natbib}

\ifthenelse{\equal{\version}{_working}}
{
}
{
\def \today {August 1, 2002}
}

\usepackage{epsfig}

\makeatletter
\newenvironment{inlinetable}{%
\def\@captype{table}%
\noindent\begin{minipage}{0.999\linewidth}\begin{center}\footnotesize}
{\end{center}\end{minipage}\smallskip}
\newenvironment{inlinefigure}{%
\def\@captype{figure}%
\noindent\begin{minipage}{0.999\linewidth}\begin{center}}
{\end{center}\end{minipage}\smallskip}
\makeatother

\shorttitle{Cluster virial masses}
\shortauthors{Irgens et al.}
\slugcomment{Accepted for publication in ApJ}
\usepackage{psfig,epsfig}
\usepackage{amsmath,latexsym}
        
\begin{document}

\title{Weak Gravitational Lensing by a Sample of \\ X-Ray Luminous
Clusters of Galaxies -- II. \\ Comparison with Virial
Masses\altaffilmark{1}}

\author{Ragnvald J.\ Irgens\altaffilmark{2} and Per B.\ Lilje}
\affil{Institute of Theoretical Astrophysics, University of Oslo, P.\
O.\ Box 1029 Blindern, \\N-0315 Oslo, Norway}
\email{per.lilje@astro.uio.no}
\author{H\aa kon Dahle\altaffilmark{3, 5}}
\affil{NORDITA, Blegdamsvej 17, DK-2100
Copenhagen \O, Denmark}
\author{S.\ J.\ Maddox\altaffilmark{4}}
\affil{School of Physics and Astronomy, University of Nottingham,
Nottingham NG7 2RD, UK}

\altaffiltext{1}{Based on
observations made with the William Herschel Telescope, operated on the
island of La Palma by the Isaac Newton Group in the Spanish
Observatorio del Roque de los Muchachos of the Instituto de
Astrof\'\i sica de Canarias, and with the Nordic Optical Telescope,
operated on the island of La Palma jointly by Denmark, Finland,
Iceland, Norway, and Sweden, in the Spanish Observatorio del Roque de
los Muchachos of the Instituto de Astrof\'\i sica de Canarias.}
\altaffiltext{2}{Present address: Telenor Broadband Services, P.\ O.\
Box 6914 St.\ Olavs plass, N-0130 Oslo, Norway}
\altaffiltext{3}{Also at: Institute for Astronomy, University of
Hawaii}  
\altaffiltext{4}{Also at: Institute of Astronomy, University of
Cambridge} 
\altaffiltext{5}{Visiting observer, University of Hawaii 2.24m Telescope
at Mauna Kea Observatory, Institute for Astronomy, University of
Hawaii}

\begin{abstract}
Dynamic velocity dispersion and mass estimates are given for a sample
of five X-ray luminous rich clusters of galaxies at intermediate
redshifts ($z\sim 0.3$) drawn from a sample of 39 clusters for which
we have obtained gravitational lens mass estimates. The velocity
dispersions are determined from between 9 and 20 redshifts measured
with the LDSS spectrograph of the William Herschel Telescope, and
virial radii are determined from imaging using the UH8K mosaic CCD
camera on the University of Hawaii 2.24m telescope.  

Including clusters with velocity dispersions taken from the
literature, we have velocity dispersion estimates for 12 clusters in
our gravitational lensing sample. For this sample we compare the
dynamical velocity dispersion estimates with our estimates of the
velocity dispersions made from gravitational lensing by fitting a
singular isothermal sphere profile to the observed tangential weak
lensing distortion as a function of radius. In all but two clusters,
we find a good agreement between the velocity dispersion estimates
based on spectroscopy and on weak lensing. 
\end{abstract}

\keywords{cosmology: observations --- dark matter --- galaxies:
clusters: individual (\objectname{Abell~914}, \objectname{Abell~959},
\objectname{Abell~1351}, \objectname{Abell~1576},
\objectname{Abell~1722}, \objectname{Abell~1995}) --- gravitational
lensing --- X-rays: galaxies: clusters}

\section{Introduction}

Rich clusters of galaxies are probably the largest well-defined and
gravitationally bound objects in the universe, and knowledge of their
properties is not only interesting in itself, but can also set
important constrains on models of large-scale structure formation. 
In particular, the cluster mass function has caught major interest
(e.g., Bahcall \& Cen 1993; Gross et al.\ 1998). Reliable estimates of
cluster masses are also important in order to constrain the ratio of
baryonic to total mass and to determine the density parameter
$\Omega_0$ (e.g., White \& Frenk 1991; Lilje 1992; White, Efstathiou,
\& Frenk 1993;
Carlberg et al.\ 1996; Eke, Cole, \& Frenk 1996).

Cluster masses have been measured using different methods, having
different biases and systematics. The method of cluster mass
measurement with the longest history is the application of the virial
theorem to positions and velocities of cluster galaxies. The first
estimates of cluster masses were obtained by \citet{Zwicky} and
\citet{Smith}. They found that the total mass of galaxies in a cluster
only accounted for a small portion of the total cluster mass. This
method works well for low redshift clusters, where it can be based on
a large number of galaxies, but has recently been applied also to
galaxies at intermediate redshifts (e.g., Carlberg et al.\ 1996) where
the number of galaxies observed per cluster is usually relatively
small.

Other methods for mass measurement are based on analysis of
gravitational lensing by the cluster and consequent distortion of
background galaxies and measurements of the X-ray temperature of the
hot intra-cluster medium. Of these, weak gravitational lensing has
recently become  frequently used for estimating the masses of clusters
at intermediate and high redshifts (see e.g., recent reviews by
Mellier [1999] and Bartelmann \& Schneider [2001]). To understand
systematic differences between the different mass estimators,
comparative studies are necessary. Dahle et al.\ (2002; hereafter
Paper I) presents weak lensing measurements of 39 highly X-ray
luminous galaxies at redshift $0.15<z<0.35$ where the masses and their
distributions have been determined homogeneously by up-to-date
methods. In this paper we present velocity dispersion and virial mass
estimates of five clusters from the sample of Paper I, which are
compared with the weak lensing results. Our mass determinations are
based on the method pioneered by \citet{LimberMathews} where the mass
estimator is
\begin{equation}
  M_V= 
  \frac{3\pi}{2}\frac{\sigma_P^2R_{H}}{G}.
  \label{eq:vmass}
\end{equation}
Here  $\sigma_P$ is the one-dimensional velocity dispersion of the
cluster, while $R_H$ is the projected mean harmonic point-wise
separation (projected virial radius) defined by
\begin{equation}
  R_H=\frac{N(N-1)}{\sum_{i<j}R_{ij}^{-1}},
  \label{eq:Rv}
\end{equation}
where $N$ is the number of galaxies in the cluster and $R_{ij}$ is the
projected distance between any pair of galaxies. In our analysis we
have used this method with the modifications of \citet{Carlberg96}
which are described later in the paper. To be able to use equation~(\ref{eq:vmass}) or versions of it for estimating masses, one has to
both determine the velocity dispersion of a cluster through
spectroscopy and the angular distribution of cluster galaxies from
photometric imaging.

In \S 2 we present our cluster sample and describe our observing and
data reduction procedures for the spectroscopic and photometric
observations of the sample.  In \S 3 we determine the velocity
dispersions of the clusters, while we in \S 4 use the photometric data
to estimate the virial radii of the clusters. The virial mass
estimates are presented in \S 5. Finally, in \S 6, we add to our
sample a sample of seven clusters for which we have obtained weak
lensing data and for which there exists velocity dispersions in the
literature. The velocity dispersions of the clusters of this larger
sample are compared with our weak lensing results.

\section{Sample and Observations}

Our clusters are taken from the weak lensing analysis sample of Paper
I, and the selection criteria are described in detail there. That
sample consists of 39 highly X-ray luminous clusters in the redshift
range $0.15 < z < 0.35$, taken from the samples of Briel \& Henry
(1993) and Ebeling et al.\ (1996, 1998). The six clusters selected for
optical spectroscopic analysis, \objectname{Abell~914},
\objectname{Abell~959}, \objectname{Abell~1351},
\objectname{Abell~1576}, \objectname{Abell~1722}, and
\objectname{Abell 1995}, were all in the \citet{Briel&Henry} sample.
Except from \objectname{Abell~914}, they were selected to have
X-ray luminosities higher than $7\times 10^{44}\,{\rm erg}\, {\rm
s}^{-1}$ in the 0.5--2.5 keV ROSAT cluster rest frame band. With the
revised redshift presented in this paper, also \objectname{Abell~959}
has a somewhat lower X-ray luminosity.  The cluster redshifts
\citep{Huchra} are in the range 0.29--0.33, except for
\objectname{Abell~914} which is at $z=0.19$. The data for
\objectname{Abell~959} were of poor quality because of clouds, and
therefore that cluster is omitted from most of our analysis.

Our spectroscopic observations were made on the three nights, 1995
April 30 to May 3, with the LDSS-2 Low Dispersion Survey Spectrograph
\citep{Allington} at the 4.2m William Herschel Telescope (WHT) on La
Palma. The LDSS-2 was used in a conventional multi-slit mode with slit
width $1\farcs 5$ and with the medium grism.  The slit length was
typically 15\arcsec -- 20\arcsec, which left adequate space for sky
line subtraction around each galaxy (the half light radii of the
galaxies were in the range 1\arcsec -- 1\farcs 5).  We used one mask
per cluster, with typically 30 to 40 slits in the $11\arcmin$
field. The detector was the thinned $1024^2$ pixel Tektronix CCD TEK-1
with an angular scale of $0\farcs 59 \,{\rm pixel}^{-1}$ giving a
dispersion of $5.3{\textrm \AA}\,{\rm pixel}^{-1}$.  The FWHM seeing
was typically 1\farcs 1 -- 1\farcs 2 during the observations. The
spectra normally covered the range 3500 to 8400\AA, for some spectra
the range was smaller.

Target galaxies were chosen based on magnitude and morphology (color
information was not used) from images taken at the Nordic Optical
Telescope (also at La Palma), coordinates were determined from the APM
Sky Catalogue\footnote{http://www.ast.cam.ac.uk/{\~\ }mike/apmcat/},
and the slit positions of the different masks were then determined in
the standard way. Each cluster mask was observed with three exposures,
each of 1.8 ks duration, giving a 5.4 ks combined exposure of each
field.  Dispersed flat-fields of the twilight sky and of a tungsten
lamp were taken for each mask. Before each cluster observation, an
exposure was made of a CuAr wavelength calibration lamp with high
signal-to-noise lines in the range 5400{\AA}--7500{\AA}. Flux
calibration exposures were made of a number of spectrophotometric
standard stars.

The data were reduced by the LEXT package (Colless et al.\ 1990;
Allington-Smith et al.\ 1994). The data reduction included bias
subtraction, correction for distortion effects, flat-fielding,
combination of exposures, sky subtraction, extraction of
one-dimensional spectra, and wavelength- and flux calibration. The
data reduction procedure took into account flexure differences between
exposures and geometric distortions in the LDSS-2 optics. The optimal
extraction routine of \citet{Horne} was utilized. The wavelength
calibration was based on third-order polynomial fits to the CuAr arc
lines, and in general provided rms residuals less than 0.2\AA\ for
wavelengths between 5400\AA\ and 8000\AA, which is the range used to
determine the redshifts. A typical reduced one-dimensional spectrum of
an early-type galaxy in one of our clusters is shown in
Figure~\ref{fig:a1576-25}.

\ifthenelse{\equal{\version}{_apj}}
{}{
\begin{inlinefigure}
\centering\epsfig{file=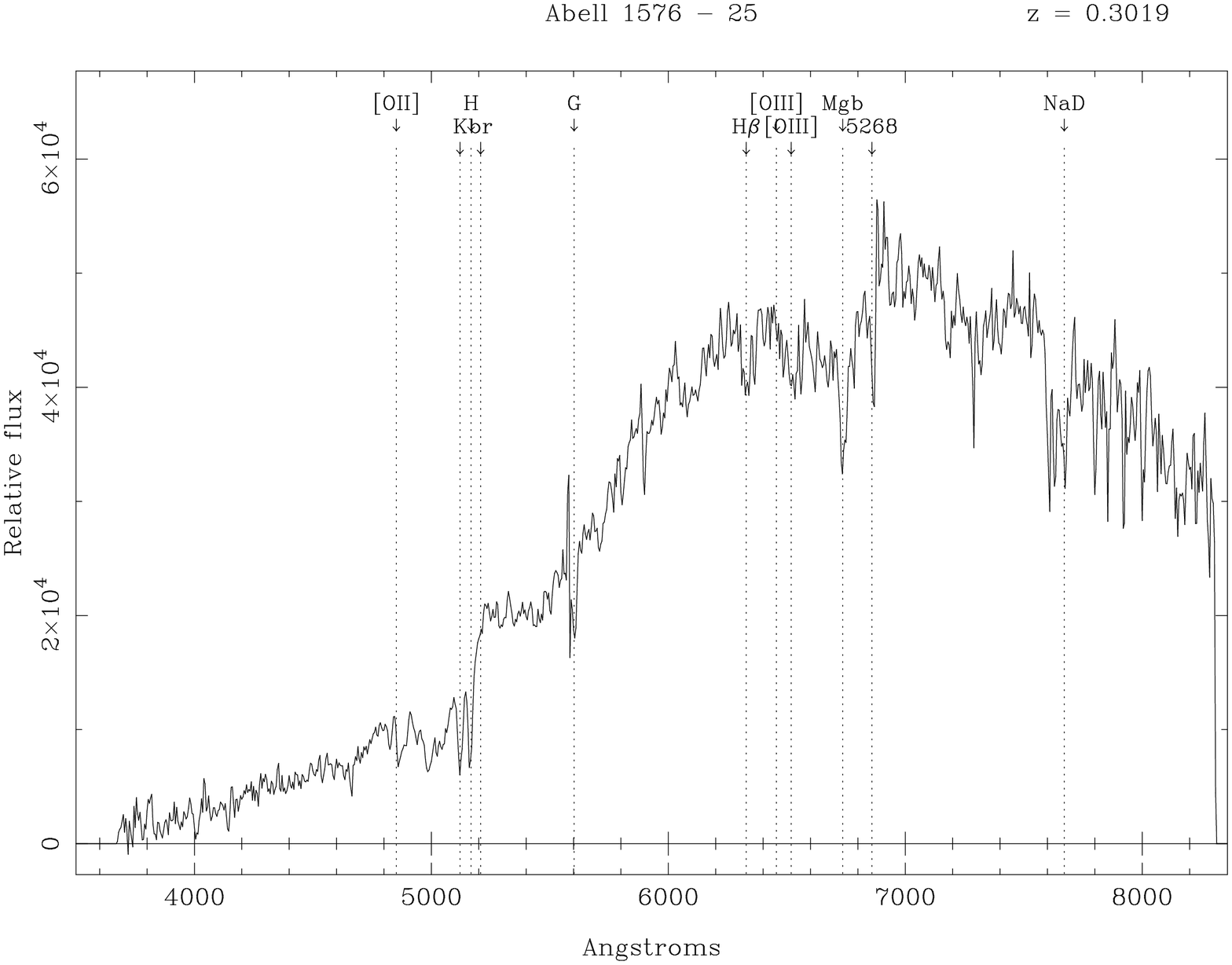,width=\figwidth}
    \caption{Spectrum of an early type galaxy in Abell~1576 with $z=0.302$.}
    \label{fig:a1576-25}
\end{inlinefigure}
}

The individual one-dimensional spectra were analyzed with the utility
{\tt crcor} which was kindly provided by Dr.\ W.\ J.\ Sutherland. Its
main estimation method is the cross correlation technique of
\citet{Tonry_Davis}, but all spectra are also searched for emission
lines. In our case, the cross correlation was done with a set of 18
template spectra, using galaxy spectra from \citet{Kennicutt} and
stellar spectra from Jacoby, Hunter, \& Christian (1984). The cross
correlation routine gives an estimate of the redshift, the
significance of the redshift and its mean error, and it also assigns a
quality flag ranging from 0 to 4. If emission lines are found, a
quality flag for the emission line redshift is also assigned based on
the number of identified lines and their strength. All spectra were
also checked manually. The cross correlation method was preferred if
it gave a quality flag equal to or better than the emission line
quality flag. In general, redshifts with quality flag 3 or better were
found to be trustworthy, and were used in the analysis. Spectra with
quality flag $\le 2$ were discarded, unless they were confirmed by
manual inspection. The data for \objectname{Abell~959}, which were
collected during partial cloud cover, only yielded five good galaxy
redshifts. For the other clusters, the number of good spectra range
from 13 (\objectname{Abell~1722}) to 24 (\objectname{Abell~1995}). The
estimated redshift error was typically 50--$100\,{\rm km}\,{\rm
s}^{-1}$. One of the objects in \objectname{Abell~1995} was not a
galaxy, but a quasar at $z=2.66$. Its coordinates (J2000.0) are
$\alpha=14^{\rm h}53^{\rm m}08.9^{\rm s}$ and $\delta=58\degr
03\arcmin 11\farcs 9$. It is located about $90\arcsec$ east of the
central galaxy of \objectname{Abell~1995}, and a simple singular
isothermal sphere gravitational lens model with a velocity dispersion
$1200\,{\rm km}\,{\rm s}^{-1}$ (which is consistent with our weak
gravitational lensing data) implies that the QSO is magnified by a
factor 1.5 by the cluster potential.

To determine the angular distribution of cluster galaxies, photometric
imaging of the cluster fields were performed at the University of
Hawaii 2.24m telescope (UH2.2m) at Mauna Kea, Hawaii during three
observing runs, 1998 February 19--23, 1999, May 14--16, and 2000 March
8--11, using the $8{\rm k}\times 8{\rm k}$ pixel UH8K mosaic CCD
camera in the f/10 Cassegrain focus. Further details of these
observations and the reduction procedures of the data are given in
Paper I.  The UH8K instrument has eight $4{\rm k}\times 2{\rm k}$ CCDs
and has a field of view of 19\arcmin\ by 19\arcmin, enough to fit not
only the central parts, but also most of the outer regions of the
clusters. However, chip \# 4 located in the NE corner was not used in
our analysis, as it had poor cosmetic quality.  The data were run
through a data processing pipeline consisting of pre-processing of the
original images, transformation from chip to celestial coordinates and
warping and averaging to produce the final images.  These final UH8K
images were rebinned into $2{\rm k}\times 2{\rm k}$, giving $0\farcs
6\, {\rm pixel}^{-1}$ resolution for the analysis reported in this
paper.

\section{Cluster redshifts and velocity dispersions}

In our case, the number of observed galaxies per cluster was modest,
ranging from 13 to 24 (and for \objectname{Abell~959} where it was not
possible to determine a velocity dispersion, five). It is obvious that
estimates of the cluster redshifts and especially velocity dispersions
(and the mass estimator is proportional to the square of the velocity
dispersion, see eq.\ [\ref{eq:vmass}]) must rely on utilizing the most
robust estimators with high resistance to avoid problems caused by
e.g., deviations from Gaussian velocity distribution and contamination
from foreground and background galaxies. In our analysis, we used the
optimal estimators for contaminated galaxy samples of this size
(Beers, Flynn, \& Gebhardt 1990), the biweight estimators of location
and scale, as estimators for the cluster redshifts and velocity
dispersions, respectively. Confidence intervals were calculated using
the bias corrected and accelerated bootstrap method \citep{Efron} with
10,000 bootstrap replications. These robust estimators and methods for
determining confidence intervals have been well tested on redshift
samples like ours (except from \objectname{Abell~959}, see e.g., Beers
et al.\ 1990; Girardi et al.\ 1993; Borgani et al.\ 1999).  For the
biweight estimator of location, we chose a value for the tuning
constant ($c = 6.0$) that gave the estimator good robustness and high
efficiency. For example, in the case of a sample approximating a
Gaussian distribution, the rejection point is $cMAD \equiv 4$ standard
deviations from the median, where $MAD$ ($= 0.6745$ in the Gaussian
case) is the median absolute deviation from the sample median (for
details, see e.g., Mosteller \& Tukey 1977).  For the biweight
estimator of scale, we chose a tuning constant $c = 9.0$ such that in
the Gaussian case, galaxies more that 6 standard deviations away from
the median would be rejected.  Our statistical analysis was performed
using the ROSTAT program kindly provided by Dr.\ T.\ C.\ Beers. A
single iteration was performed for the biweight measurements.

Our redshift and velocity dispersion estimates are presented in
Table~\ref{tab:location}. The analysis uses only galaxies with well
measured redshifts which have not been rejected as foreground or
background objects. The second column in Table~\ref{tab:location},
$N$, shows the number of galaxy redshifts used in each cluster. The
third and fourth columns show the derived cluster redshift and the
derived one-dimensional velocity dispersion together with 68\%
confidence limits. Given the modest number of spectroscopically
measured galaxies in each cluster, we performed an additional estimate
of the errors associated with small-number statistics to check if they
were compatible with the confidence intervals in
Table~\ref{tab:location}.  This was done by repeatedly drawing random
samples of galaxies (with $N = 10$ and $N = 20$) from a much larger
redshift data set ($N \simeq 200$; see Yee et al.\ 1996) for
\objectname{Abell~2390} ($z = 0.23$), which has a velocity dispersion
of $\sigma_P = 1093\pm 61\,{\rm km}\, {\rm s}^{-1}$ (Carlberg et al.\
1996).  The variance in the values for the redshift and velocity
dispersion calculated from $100$ such subsamples is similar to, or
slightly ($< 30\%$) larger than, the confidence intervals in
Table~\ref{tab:location}.

\ifthenelse{\equal{\version}{_apj}}
{}{
\begin{table*}
\caption{Cluster redshifts and velocity dispersions}
\centering
\begin{tabular}{cccc|cc}
\tableline
\tableline
  & $N$ & $z$ & $\sigma_P$ (${\rm km}\,{\rm s}^{-1} $) & $z$ (H90)\tablenotemark{a} &  $N$ (H90)\tablenotemark{b} \\
\tableline 
  \objectname{Abell~914} & 11 & $0.1934^{+0.0017}_{-0.0012}$ &  $1140^{+290}_{-170}$ & 0.1941 & 3 \cr
  \objectname{Abell~959} & 4\tablenotemark{c} &  $0.2857^{+0.0034}_{-0.0034}$ & \nodata & 0.3530 & 2\cr
  \objectname{Abell~1351} & 17 & $0.3279^{+0.0014}_{-0.0015}$ & $1680^{+340}_{-230}$ & 0.3220& 3 \cr
  \objectname{Abell~1576}& 14   & $0.2986^{+0.0011}_{-0.0013}$ & $1040^{+170}_{-120}$ & 0.3020& 1 \cr
  \objectname{Abell~1722}& 9 & $0.3264^{+0.0013}_{-0.0017}$ & $970^{+280}_{-130}$ & 0.3275& 1 \cr
  \objectname{Abell~1995} & 20 & $0.3207^{+0.0010}_{-0.0010}$ & $1130^{+150}_{-110}$ & 0.3180 & 2 \cr
  \tableline
\end{tabular}
  \tablenotetext{a}{Literature redshift \citep{Huchra}.}
  \tablenotetext{b}{Number of individual galaxy redshifts measured by
    \citet{Huchra}.}
  \tablenotetext{c}{With only four cluster galaxies with
    measured redshifts in the \objectname{Abell~959} field, a velocity dispersion could not be estimated. The confidence interval for the
  cluster redshift was estimated from the variance in the results obtained by randomly drawing redshift values from a much larger data set for a similar cluster (see the text for details).}
\label{tab:location}
\end{table*} 
}

\ifthenelse{\equal{\version}{_apj}}
{}{
\begin{inlinefigure}
\centering\epsfig{file=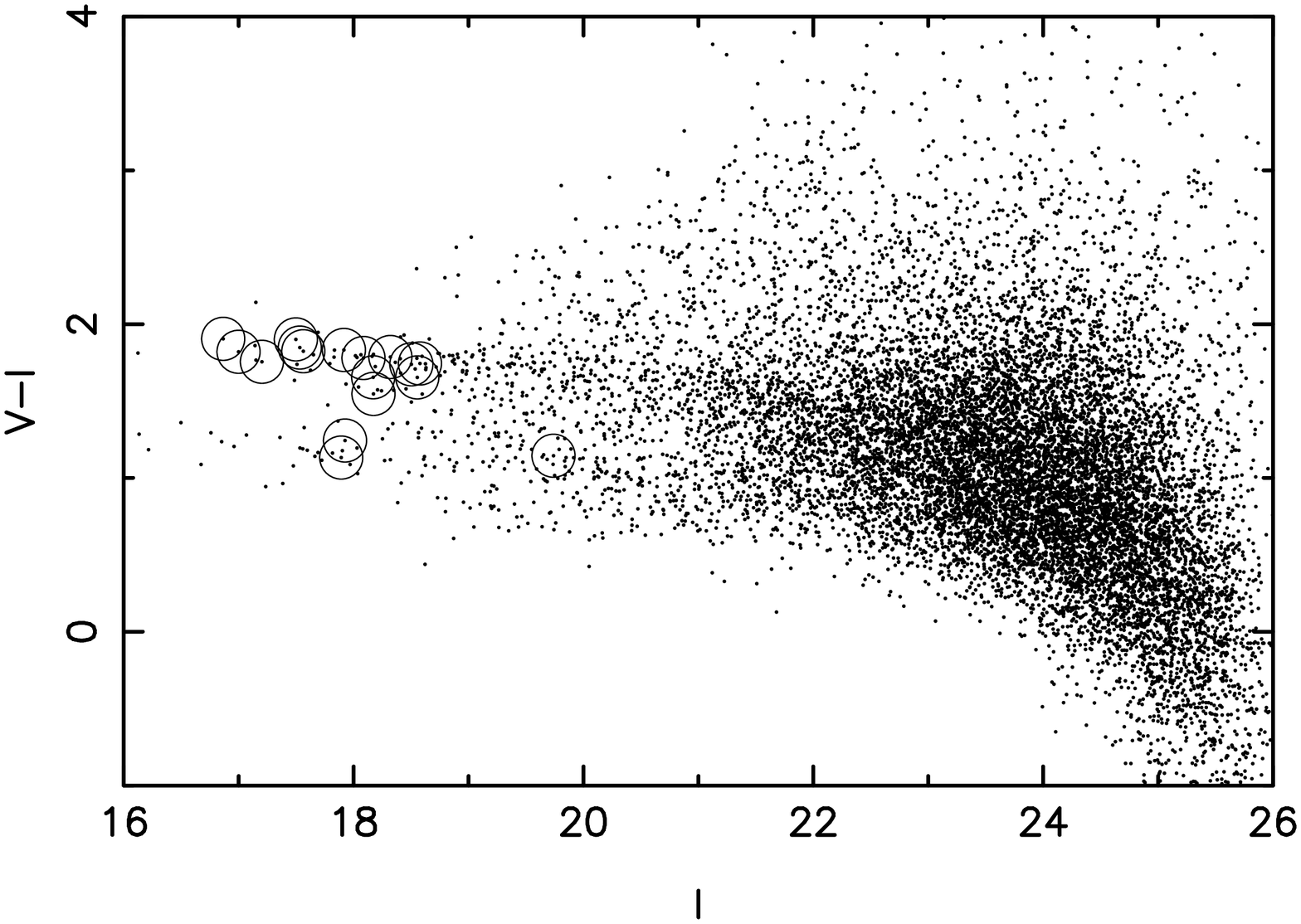,width=\figwidth}
\caption{Color-magnitude diagram for Abell~1351. The positions of the
    galaxies with measured redshifts and colors are plotted with open
    circles. With a few exceptions, the selected galaxies are all
    within the main cluster sequence.}
\label{fig:a1351colmag}
\end{inlinefigure}
}

For most of our six clusters, the determined redshifts are in
relatively good agreement with previous results \citep{Huchra}, but
ours are based on a much larger number of galaxy redshifts.  However,
for \objectname{Abell~959} where we only base our cluster redshift on
four galaxies, our redshift is significantly lower than the redshift
of \citet{Huchra}. Three of the four galaxies that we based our
redshift estimate on have $V-I$ colors putting them in the middle of
the red sequence of the color-magnitude diagram of the field. Both the
galaxies used by \citet{Huchra} were substantially bluer than the
cluster red sequence. We therefore believe that our cluster redshift
for \objectname{Abell~959} is correct.  The 68\% confidence limits for
the redshift of this cluster could not be reliably estimated using the
bootstrap method we used for the other clusters, and the confidence
limits shown in Table~{\ref{tab:location} are instead based on the
variance in the results obtained by drawing 4 galaxies at random from
the Yee et al.\ (1996) data set for \objectname{Abell~2390}.

Struble \& Rood (1999) list a redshift of $z = 0.279$ for
\objectname{Abell~1576}, significantly lower than the value we derive
from our data and the value measured by \citet{Huchra}.  The lower
redshift value (which is also the one given by NED) originates from a
study by Leir \& van den Bergh (1977) and is an estimate based on a
combination of cluster galaxy photometry and cluster richness.  We
note that \objectname{Abell~1351}, where the measurement is based on
17 redshifts, has an abnormally high velocity dispersion. This is
partially confirmed by the weak lensing analysis (see \S 6). Patel et
al.\ (2000) have recently published redshift and velocity dispersion
measurements of \objectname{Abell~1995}, based on the redshifts of 15
galaxies, six of which are common with our data set.  Their result of
$z=0.322\pm 0.001$ and $\sigma_P=1282^{+153}_{-120}\,{\rm km}\, {\rm
s}^{-1}$ agrees quite well with our result.

\section{Angular distribution of cluster galaxies}

As described in \S 1, the virial mass estimator (eq.~\ref{eq:vmass})
traditionally uses the projected mean harmonic point-wise separation
defined in equation (\ref{eq:Rv}). However, this estimator is
sensitive to close pairs and is somewhat noisy (see e.g., Carlberg et
al.\ 1996). It will also systematically underestimate the radius for a
rectangular aperture when the entire cluster is not covered (Bahcall
\& Tremaine 1996). \citet{Carlberg96} proposed to instead use a
related radius, the ring-wise projected harmonic mean radius, defined
by
\begin{gather}
  \begin{split}
    R_h&= \frac{N(N-1)}{\sum_{i<j}\frac{1}{2\pi}\int_0^{2\pi}
    \frac{d\theta}{\sqrt{R_i^2+R_j^2+2R_i R_j\cos\theta}}}
    \\
    &= \frac{N(N-1)}{\sum_{i<j}\frac{2}{\pi(R_i+R_j)}K(k_{ij})},
  \end{split}
  \label{eq:rphmr}
\end{gather}
where $R_i$ and $R_j$ are the distances of galaxies $i$ and $j$ from
the cluster center, $k_{ij}^2=4R_iR_j/(R_i+R_j)^2$ and $K(k)$ is the
complete elliptic integral of the first kind in Legendre's notation.
This radius treats one of the particles in the pairwise potential
$|\mathbf{R}_i-\mathbf{R}_j|^{-1}$ as having its mass distributed like
a ring around some given cluster center and is less noisy and handles
non-circular apertures better than the standard $R_H$. It will,
however, overestimate the true projected virial radius if the cluster
is significantly flattened or extensive sub-clustering is present. In
our analysis we therefore use both methods to check the influence of
the choice of radius in the virial mass estimator on the uncertainty
of the resulting virial mass.
 
The radii $R_h$ and $R_H$ should in principle be computed for all
galaxies residing in the virialized regions of each cluster.  In our
case, we have only measured redshifts of a few galaxies in each field,
and redshifts can therefore not be used to select likely cluster
galaxies over which these sums are to be taken. Instead we use the
UH8K photometry described in \S 2 to select the cluster galaxies.  Our
procedures for object detection and classification follow Kaiser,
Squires, \& Broadhurst (1995), and are described in detail in Paper
I. Briefly, we first ran a peak finding algorithm with a $4\sigma$
detection limit. Then we performed aperture photometry on the detected
objects and removed stellar objects brighter than $I=21^m$ by applying
cuts in the size-magnitude diagram. After having removed stars and
very faint objects (the completeness limit is $I\sim 25.0^m$), the
$V$- and $I$-catalogs were combined. In the final catalog we kept
objects that were detected in both $V$ and $I$ with positions that
differed by less than $1\farcs 2$ (2 pixels). The peak finding
algorithm of Kaiser et al.\ (1995) is optimised for detecting small,
faint galaxies, and may occasionally have problems resolving closely
grouped galaxies in cluster cores. To test whether our choice of
galaxy detection algorithm had a significant influence on our results,
we also calculated $R_H$ values from galaxy catalogs generated using
the SExtractor object finding algorithm (Bertin \& Arnouts 1996). The
resulting variations in the $R_H$ values were small and within the
quoted uncertainty intervals.  In addition, the performance of the
Kaiser et al.\ (1995) detection algorithm was tested by manually
examining images of cluster fields with the catalog object positions
superposed as small circles. The positions indicated by the Kaiser et
al.\ algorithm were indeed almost always centered on the brightest
galaxy within the circle (for an example, see
Fig.~\ref{fig:a1351members}), and we could not manually find obvious
cluster members not found by the algorithm. 

\ifthenelse{\equal{\version}{_apj}}
{}{
\begin{figure*}
\centering\epsfig{file=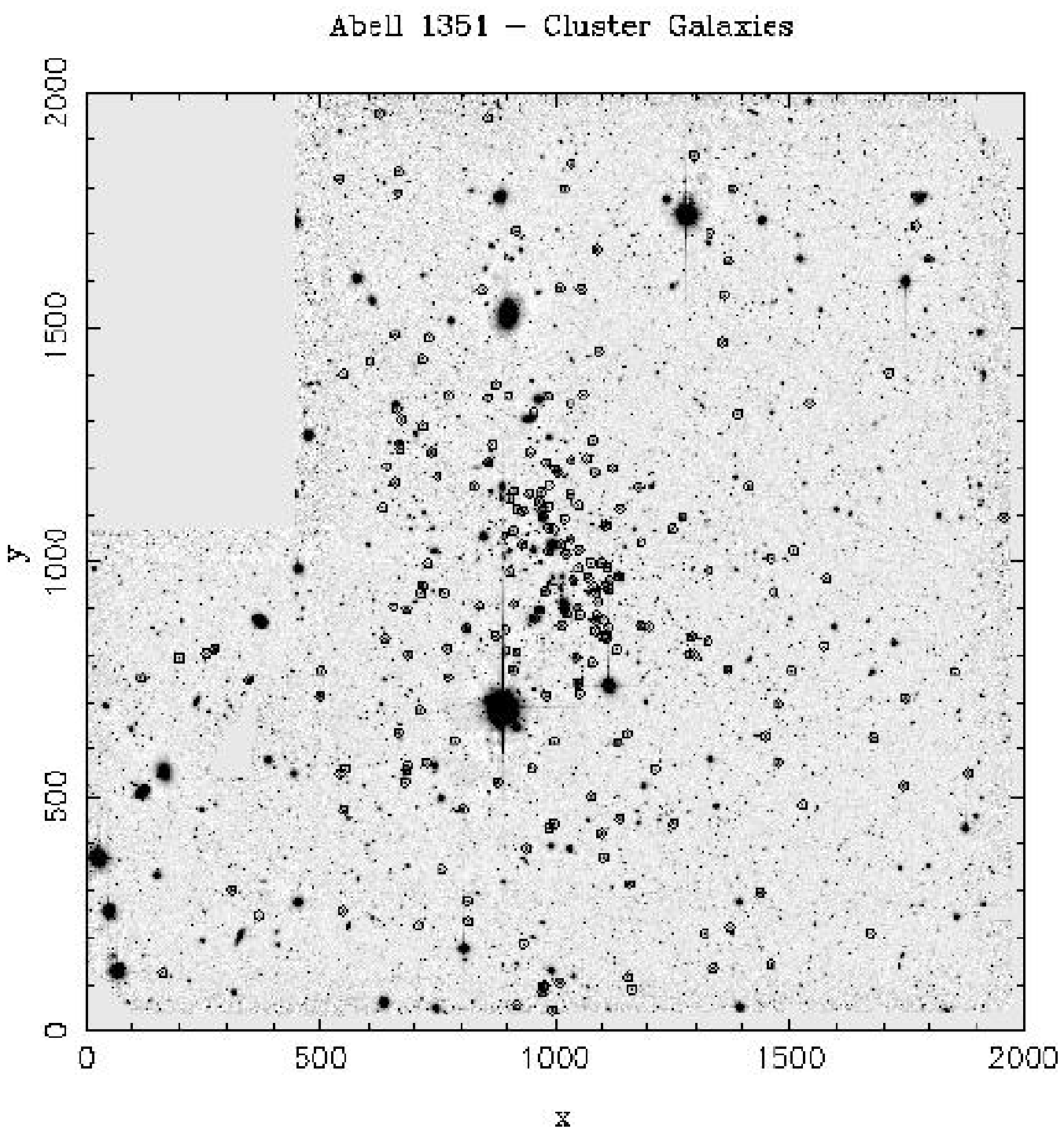,width=\figwidth}
\caption{Image of Abell~1351 overlaid positions of cluster
galaxies identified for the color-magnitude diagram (circles). In
this figure, north is up, east is left and the scale on the axes
is in pixels of size 0\farcs 6.}
\label{fig:a1351members}
\end{figure*}
}

In a color-magnitude diagram (see Fig.~\ref{fig:a1351colmag} for
\objectname{Abell~1351}), the cluster galaxy red sequence consisting
of early type galaxies is clearly visible (see also e.g., Olsen
[2000]).  We exploited this fact in our procedures for discriminating
cluster members from galaxies possibly belonging to background or
foreground structures.  As described in detail in Paper I, we
generated catalogs of ``cluster'' galaxies by making appropriate cuts
in each cluster color-magnitude diagram to include early type galaxies
at the cluster redshift.  We also tried to use galaxy samples that
included all galaxies in the fields down to a certain magnitude limit
without any color selection. This did not produce significant changes
in the results, except for \objectname{Abell~1351}, which displays
clear evidence of a foreground structure (the color-magnitude diagram
for this cluster clearly shows an additional red sequence at $V-I \sim
1.2$).  To assess the uncertainty in the analysis caused by some
arbitrariness in the exact positions of the color cuts, we varied the
limits of the cuts to make an ensemble of galaxy catalogs for each
cluster on which the further analysis was performed in parallel. To
determine $R_h$, the position of the cluster center is needed. This
was found in two steps, first by applying a peak finding algorithm on
a smoothed map of the distribution of galaxies defined to be cluster
members by the color criteria described above. As described later in
this chapter, we chose three different sets of galaxies within three
different radii from this center to perform our analysis. For each of
these sets we recomputed the center which we used for determining
$R_h$. The center was set to be the biweight estimate (Beers et al.\
1990) of the mean of the $x$ and $y$ coordinates of all galaxies used
in the estimate. We tested for the effect of varying the center
position, and found that varying this by less than 50 pixels did not
significantly affect the results.

The radius entering the virial mass estimator (eq.~\ref{eq:vmass}),
and hence the mass estimate itself, can be biased if we include
galaxies outside the virialized part of the cluster, or vice versa, if
we exclude galaxies outside some radius within the virialized part of
the cluster. The projected density distribution of galaxies outside
the cluster core is well known to fall approximately $\propto R^{-1}$
both in the virialized region and in the surrounding infalling region
(reflected in the cluster-galaxy cross-correlation function [e.g.,
Lilje \& Efstathiou 1988]). Thus we would not expect to see a visible
border between the virialized region and the infalling region. This is
confirmed by our data. However, it is normally assumed (see e.g.,
Carlberg et al.\ 1996) that the virialized region encompasses the
region around the cluster center where the average density is about
200 times the critical density of the universe. This is based on the
fact that in the Einstein de Sitter universe, a purely spherical
top-hat perturbation will virialize at a density 178 times the
background density at the time of virialization. In other universe
models, the value is somewhat different, and in open models all
clusters virialize at an epoch much earlier than the present
epoch. However, the figure 200 is sufficient as an approximate
estimate (see e.g., Lilje 1992).  For a typical rich cluster this is
the region within a radius quite close to the Abell radius
($1.5h^{-1}{\rm Mpc}$). Our 19\arcmin\ by 19\arcmin\ field is about
two Abell radii across at the typical cluster redshifts, so our field
is approximately of the right size to minimize this possible bias. In
our analysis, we compute $R_H$ and $R_h$ both using all ``cluster''
galaxies in the field, and for only the galaxies within radii of 6
respectively 8 arcminutes from the cluster center. The scatter between
these three different estimates of the radii estimated from the same
cluster was normally of the order of 10\%. There was a systematic
effect, in that the estimated radii to some extent scaled with the
size of the cutoff.  We use the differences in these results to
determine the uncertainty in our final virial mass determinations. As
seen in Figure~\ref{fig:a1351members} where the positions of all
``cluster galaxies'' in a typical catalog for \objectname{Abell~1351}
are shown as circles, most of the galaxies defined as ``cluster''
galaxies in each field do indeed reside within the inner of these
radii.  Before doing our analysis we also had to mask certain areas in
the fields and reject galaxies in those areas from our analysis. Those
areas included the UH8K CCD \# 4 and the areas surrounding bright
stars. For \objectname{Abell~914}, the cluster was not positioned in
the center of the field but in the NW quadrant, to be able to
simultaneously image another cluster in the SE quadrant. For
\objectname{Abell~914} we therefore only used galaxies in the NW
$15\arcmin\times 15\arcmin$ part of the mosaic in our analysis.

For each cluster we constructed three ensembles of ``cluster'' galaxy
catalogs, using the whole image, the central 8\arcmin\ field and the
central 6\arcmin\ field. Each ensemble consisted of about 5 catalogs
where the color and magnitude selection criteria for ``cluster''
galaxies had been varied within reasonable bounds. For each catalog in
all these ensembles we computed $R_H$ (eq.~\ref{eq:Rv}) and $R_h$
(eq.~\ref{eq:rphmr}). For all the ensembles, $R_h$ differed from $R_H$
by less than 5\%, and we chose to only use $R_h$ in our further
analysis. We express $R_h$ in length units assuming an Einstein
de~Sitter cosmology ($\Omega_0=1$, $\Lambda=0$). If we instead had
assumed a cosmology with $\Omega_0=0.3$ and $\Lambda=0.7$, the radii
would have become 13\% larger at $z=0.3$ and 9\% larger at
$z=0.2$. Compared to our other uncertainties, this error is relatively
small. Finally, we express the radius used in our further virial mass
analysis as the deprojected virial radius,
\begin{equation}
r_V=\frac{\pi}{2}R_h.
\end{equation}

Dr.\ T.\ C.\ Beers' ROSTAT program (described in \S 3) was then
applied to the combined ensemble of estimates of $r_V$ for each
cluster to find the best estimate of its $r_V$ (by the biweight
estimator of location), and the 68\% confidence limits on this
estimate (by the bias corrected and accelerated bootstrap
method). With the systematic effect mentioned above, i.e., a
correlation between the estimated $R_h$ and the radius within which
galaxies have been picked for analysis, it is clear that this method
has its deficiencies. It does not necessarily give an unbiased
estimate of the expectancy value of $r_V$, and the estimated
confidence limits cannot be treated as formal 68\% confidence
limits. However, with the rather large uncertainties in these
estimates, the quoted confidence limits give a fair estimate of the
uncertainty in the results, which are of order 10\%. Our results are
presented in the second column of Table~\ref{tab:radii1}. Since the
small number of redshifts for \objectname{Abell~959} prevented us from
finding its velocity dispersion, we also omitted it from the galaxy
distribution analysis.

\ifthenelse{\equal{\version}{_apj}}
{}{
\begin{inlinetable}
\caption{Cluster Virial Radii and Masses}
\begin{tabular}{ccc}
\tableline
\tableline
 & $r_V$ ($h^{-1}{\rm Mpc}$) & $M_V$ ($10^{15} h^{-1}M_\odot$) \\
\tableline
\objectname{Abell~914} & $1.23^{+0.13}_{-0.12}$&$1.1 \pm 0.6$ \cr \objectname{Abell~1351} & $1.68^{+0.18}_{-0.13}$&$3.3 \pm 1.4$
\cr \objectname{Abell~1576} & $1.52^{+0.10}_{-0.21}$ &$4.0 \pm 1.4$ \cr
\objectname{Abell~1722} & $1.86^{+0.22}_{-0.19}$ &$1.2\pm 0.6$ \cr
\objectname{Abell~1995} & $1.19^{+0.15}_{-0.20}$&$1.1 \pm 0.4$ \cr
\tableline
\end{tabular}
\tablecomments{The values are given for an Einstein-de Sitter
Universe. See the text for a brief discussion of possible additional
systematic uncertainties that are not included in the error intervals
given here.}
\label{tab:radii1}
\end{inlinetable}
}

\section{Cluster virial masses}
\label{chap:vmass}

The virial masses of the clusters, given by equation~(\ref{eq:vmass}),
are shown in the third column of Table~\ref{tab:radii1}. The given
statistical errors are based on half the length of the 68\% confidence
intervals of the cluster velocity dispersions and virial radii, and
standard propagation of errors.

The virial mass given by equation~(\ref{eq:vmass}) and
Table~\ref{tab:radii1} is the mass of the virialized region of the
cluster, supposed to be within the virial radius $r_V$. Since the
density profile in the outer parts of the virialized region, and at
least the inner part of the infalling region, in a cluster is close to
being $\propto r^{-\alpha}$ where $2\lesssim\alpha\lesssim 3$ (e.g.,
Adami et al.\ 1998; Carlberg et al.\ 1996; Lilje \& Efstathiou 1988),
this mass can be scaled to other radii assuming that the mass inside a
given radius grows approximately proportional to the radius.

As pointed out in \S 4, there is no clear feature seen in clusters at
the transition from the virialized region to the surrounding infalling
region. Also, the idea of a virialized region bounded in space by a
spherical border, surrounded by the non-virialized infalling region,
is clearly an oversimplification. There might well be recently
infallen galaxies or subclusters within the virial radius which have
not been virialized, and these will bias the mass estimate. Also, we
know that real clusters may be far from spherical. Even the quite
large statistical errors given in Table~\ref{tab:radii1}, should
therefore be taken with a pinch of salt, because of the uncertainty in
what the derived virialized mass really measures.

Even for our idealized model, there are also several other possible
sources for systematic errors. If our measurement only uses galaxies
inside a bounding radius $r_b$ much smaller than the virial radius,
our mass estimate will be an overestimate of the mass within $r_b$
because of the neglect of the surface term in the virial theorem
\citep{Carlberg96}. This effect is at most 50\%. Our virial radius
estimates of \S 4 were derived assuming an Einstein-de Sitter
cosmology. As pointed out in that chapter, this overestimates the
virial radius, and hence the virial mass, by about 10\% if the real
universe is vacuum energy dominated with $\Omega_0\sim 0.3$.

\section{Comparison with weak lensing measurements} 
\label{Chap:comp}

In Paper I we derived weak lensing mass estimates for a sample of 39
galaxy clusters including the clusters studied in this paper.  Seven
additional clusters from the Paper I sample have spectroscopically
determined velocity dispersion measurements available in the
literature.  Here, we compare these measurements to the velocity
dispersion estimates $\sigma_{\rm WL}$ derived in Paper I by fitting a
singular isothermal sphere (SIS) profile to the observed tangential
weak lensing distortion as a function of radius. We note, however,
that three of the clusters (\objectname{Abell~959},
\objectname{Abell~1351} and \objectname{Abell~1576}) were poorly fit
by a SIS profile.

In Table~\ref{tab:cluster_data} we summarize the available data for
the clusters discussed in this paper, and the available data for the
clusters with published velocity dispersion measurements for which we
have weak lensing data.

The quantity $\sigma_{\rm DM}$ (= $\sigma_{\rm WL}$ listed in Paper I)
in the fourth column of Table~\ref{tab:cluster_data} is derived from
the weak lensing results of Paper I. A subsample of 11 clusters have
spectroscopically measured velocity dispersions $\sigma_P$, which are
plotted against the dark matter velocity dispersions $\sigma_{\rm DM}$
in Figure~\ref{fig:sigmav}.

\ifthenelse{\equal{\version}{_apj}}
{}{
\begin{table*}
\caption{Cluster data.}
\centering
\begin{tabular}{ccrcrc}
\tableline
\tableline
Designation &
  $z$& $\sigma_{P}$ & Ref. &
  $\sigma_{\rm DM}$ & $L_{\rm X}$ \cr &   &   &   &
    & (0.1--2.4 keV) \cr   &  & (${\rm km}\,{\rm s}^{-1}$)
  & $\sigma_{P}$ & (${\rm km}\,{\rm s}^{-1}$) &
  $10^{44}$~${\rm erg}\,{\rm s}^{-1}$ \cr 
\tableline
  Abell 115 & 0.197 &
  $1074^{+208}_{-121}$ & (1) & $1130^{+210}_{-270}$&
  14.59\tablenotemark{a} \cr Abell 520 & 0.203& $988 \pm
  76$ & (2) & $1050 \pm 100$ & 14.52\tablenotemark{b} \cr Abell 665 & 0.182 & $821^{+233}_{-130}$ & (1) &
  $1010^{+150}_{-170}$ & 15.69\tablenotemark{b} \cr Abell 697 & 0.282& $941 \pm 296$ & (3) & $1730^{+190}_{-200}$ &
  19.15\tablenotemark{b} \cr Abell 914 & 0.193 &
  $1141^{+153}_{-120}$ & \nodata & $540^{+160}_{-190}$ &
  $5.00$\tablenotemark{b} \cr Abell
  959\tablenotemark{c} & 0.286 & \nodata & \nodata &
  $990^{+100}_{-110}$ & $14.3$\tablenotemark{b} \cr Abell 963 & 0.206 &$1350^{+200}_{-150}$ & (4) &
  $1070^{+150}_{-160}$ & 8.72\tablenotemark{b} \cr Abell 1351 & 0.328 &$1680^{+340}_{-229}$ & \nodata &
  $1410^{+80}_{-90}$ & $8.31$\tablenotemark{b} \cr
  Abell 1576 & 0.299 &$1041^{+153}_{-183}$ & \nodata & $1060 \pm 90$ &
  $11.52$\tablenotemark{b} \cr Abell 1722
  & 0.326 &$966^{+283}_{-132}$ & \nodata & $1160 \pm 140$ &
  $9.70$\tablenotemark{b} \cr Abell
  1995\tablenotemark{d} & 0.321 & $1126^{+151}_{-105}$ & \nodata &
  $1240 \pm 80$ & $13.42$\tablenotemark{b} \cr Abell 2104 & 0.153 & $1200 \pm 200$ & (5) & $1390\pm 180$ &
  7.89\tablenotemark{e} \cr Zwicky 7160 & 0.258
  &$1133 \pm 140$ & (2) & $1230 \pm 130$ & 13.73\tablenotemark{b} \cr 
\tableline 
\end{tabular} 
\tablerefs{ (1) Girardi \& Mezzetti
  2001, (2) Carlberg et al.\ 1996, (3) Metzger \& Ma 2000, (4) Lavery
  \& Henry 1998, (5) Liang et al.\ 2000} \tablenotetext{a}{X-ray luminosity measured by
  Ebeling et al.\ (1998).}  \tablenotetext{b}{X-ray luminosity measured
  by B{\"o}hringer et al.\ (2000).}  \tablenotetext{c}{The values have
  been corrected for the revised redshift for Abell 959 presented in
  this paper.}  \tablenotetext{d}{Patel et al.\ (2000) measure
  $\sigma_P = 1282^{+153}_{-120}$ for Abell 1995.}
  \tablenotetext{e}{X-ray luminosity measured by Ebeling et al.\ (1996).}
\label{tab:cluster_data}
\end{table*}
}

There is generally a reasonably good agreement between $\sigma_P$ and
$\sigma_{\rm DM}$, with a tendency for the former to be smaller than
the latter by a barely significant amount, the median value of
$\sigma_P / \sigma_{\rm DM}$ being 0.94. We also find $\langle
\sigma_P / \sigma_{\rm DM} \rangle = 1.02 \pm 0.13$ when using all
twelve clusters and $\langle \sigma_P / \sigma_{\rm DM} \rangle = 0.97
\pm 0.08$ after excluding two outliers mentioned below. If we only
consider the clusters for which we have measured $\sigma_P$ in this
paper (but excluding \objectname{Abell~914}, for reasons given below),
we find $\langle \sigma_P / \sigma_{\rm DM} \rangle = 0.98 \pm 0.12$.

One cluster (\objectname{Abell~914}, at $z = 0.194$) show a $\sigma_P$
a factor $\sim 2$ larger than $\sigma_{DM}$, with a very strong ($>
2\sigma$) discrepancy between the two velocity dispersion
values. However, there is a second cluster (\objectname{Abell~922}, at
$z = 0.189$) at a projected distance of only 11 arcminutes on the
sky. Since the weak lensing measurements indicate that
\objectname{Abell~922} is a more massive cluster than
\objectname{Abell~914} (its estimated $\sigma_{\rm DM}$ is
$810^{+120}_{-130}\,{\rm km}\,{\rm s}^{-1}$), the observed redshifts,
which were measured in a $10' \times 10'$ field centered on
\objectname{Abell 914}, may have been significantly contaminated by
galaxies along the line of sight belonging to the other cluster.

A second cluster (\objectname{Abell 697}, at $z = 0.282$) shows an
almost equally strong discrepancy in the opposite direction, with
$\sigma_{DM} \sim 2 \times \sigma_P$. We note however, that the
spectroscopically measured galaxy velocity dispersion is based on a
rather small number of cluster galaxies ($N = 9$; see Metzger \& Ma
2000).

\ifthenelse{\equal{\version}{_apj}}
{}{
\begin{inlinefigure}
\centering\epsfig{file=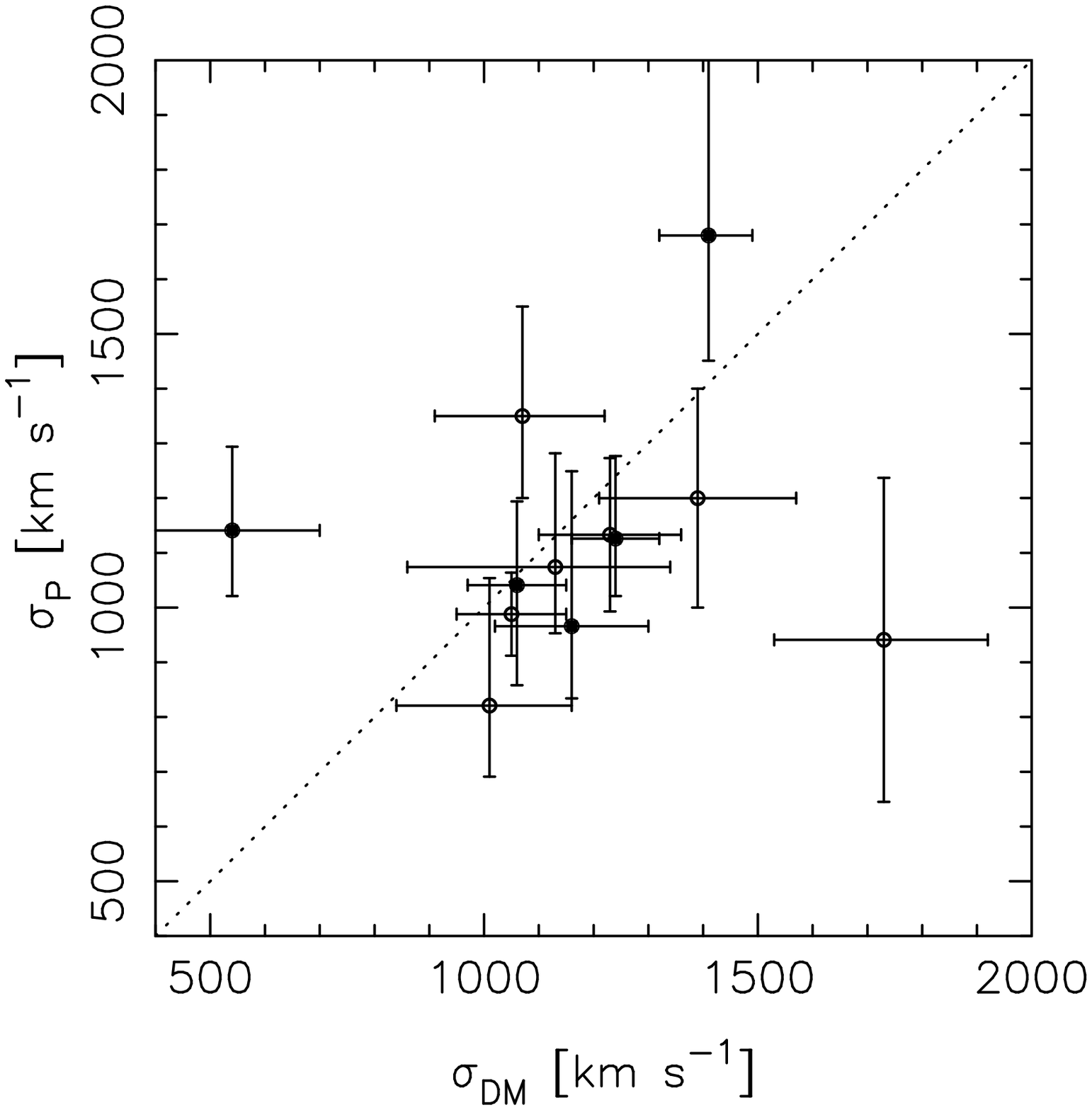,width=\figwidth}
\caption[Galaxy vs.\ dark matter velocity dispersion]{The
  spectroscopically measured velocity dispersion $\sigma_P$ vs.\ the
  dark matter velocity dispersion $\sigma_{\rm DM}$. The error bars
  shown are at $1\sigma$. The dotted line indicates equality between
  the values. Filled circles represent the clusters for which we present 
  $\sigma_P$ measurements in this paper. The other points have $\sigma_P$ 
  values drawn from the literature.}
\label{fig:sigmav}
\end{inlinefigure}
}

\section{Discussion}
\label{chap:discussion}

Our in general good agreement between $\sigma_P$ determined from
dynamical estimates and $\sigma_{DM}$ determined from gravitational
lensing estimates contrast other results. 

Smail et al.\ (1997) presented a comparison of weak lensing and
spectroscopic mass measurements for a sample of 9 clusters at $0.17 <
z < 0.55$ which were imaged with the WFPC2 instrument aboard the
Hubble Space Telescope (HST).  They compared the average tangential
shear observed within an annulus around the optical cluster center
($60 < r < 200 h^{-1}\,{\rm kpc}$) to the expected shear from SIS-type
mass profile with the observed galaxy velocity dispersion. The shear
signal was found to be significantly lower than predicted from the
spectroscopic measurements for all the clusters, and Smail et al.\
(1997) suggested that this was caused by systematic overestimates (by
typically $\sim 50$\%) of the cluster velocity dispersion values given
in their paper. As noted in \S\ref{Chap:comp}, we see no such
discrepancy in our data.  Hoekstra et al.\ (2001) argue that this
discrepancy is instead a result of the small field size of the single
WFPC2 pointings used by Smail et al.\ (1997) to measure the
gravitational shear. If there is significant substructure in the
central regions of the clusters, or if the density profiles of the
clusters are significantly shallower than a SIS-type model at small
radii, the shear measurements are likely to underestimate the true
cluster mass.

Brainerd et al.\ (1999) have used ray-tracing through high-resolution
$N$-body simulations of massive ($M_V \sim 10^{15}h^{-1}M_{\sun}$)
clusters to show that the average tangential shear, when measured
within an annulus with an outer radius $R_{\rm MAX}$ significantly
smaller than the virial radius (and assuming a SIS-type mass profile),
will tend to underestimate the true cluster mass. For $R_{\rm MAX}
\simeq R_V$, they find that the SIS-model fit yields quite accurate
mass values (typically 5\% -- 10\% underestimates; see their Figure~9)
even if the true mass density profile of the clusters is more
accurately represented by the model proposed by Navarro, Frenk, \&
White (1995). We note that for most of the clusters listed in
Table~\ref{tab:cluster_data}, we have estimated $\sigma_{\rm DM}$
based on tangential shear measurements out to radii $\simeq R_V$ (for
most of our clusters, we have used $R_{\rm MAX} = 550\arcsec = 1.5
h^{-1}\,{\rm Mpc}$ at $z = 0.3$; see Paper I for details), so we
expect our values to be essentially unbiased.
  
Girardi \& Mezzetti (2001) have reanalyzed the available radial
velocity measurements in the clusters studied by Smail et al.\ (1997)
and derive velocity dispersion values that are on average about 20\%
lower than those given by Smail et al.\ (1997). Thus, it appears that
the discrepancy between spectroscopic and lensing mass estimates seen
by Smail et al.\ (1997) is most probably caused by a simultaneous
overestimation of dynamical masses and underestimation of lensing
masses, of almost equal magnitude.  Girardi \& Mezzetti (2001) also
compare their virial cluster mass estimates with weak lensing mass
estimates drawn from the literature for 18 clusters (for 7 of these
the lensing mass was measured from single HST WFPC2 pointings), and
find a median value $M_{spec}/M_{lens} = 1.30$ and a 90\% confidence
interval $0.63 < M_{spec}/M_{lens} < 2.13$. This is consistent with
our results in \S\ref{Chap:comp} within the errors, but the $M_{lens}$
values they use are probably subject to a larger bias caused by the
(on average) significantly smaller fields used to measured the shear.

In their study, Brainerd et al.\ (1999) did not consider the possible
effect of structures outside their clusters on the lensing mass
estimates.  Galaxy clusters are believed to be connected by large
filaments of dark matter, diffuse gas, and galaxies, and mass within
such filaments will contribute significantly to the lensing mass when
seen in projection along the line of sight to a cluster, even if it is
physically separated from the cluster by several Mpc (Metzler, White,
\& Loken 2001).  Still, even in this scenario, $\sigma_{\rm DM}$ may
not necessarily be significantly larger than $\sigma_P$, since the
galaxy velocity dispersion would also be biased upwards by galaxies
outside the virial radius falling into the main cluster along the
filaments.

The modest fraction ($\sim 30$\%) of massive clusters that have
significant substructure in their dark matter distribution provide
evidence that the majority of massive clusters at $z = 0.2$--0.3 are
close to dynamical equilibrium (Dahle et al.\ 2002a).  The good
agreement that we find between virial masses and lensing masses for
most of our clusters is consistent with this picture, although
significant amounts of substructure could probably be present in the
clusters without changing the $\sigma_P$ values beyond the error
bounds of our measurements. To reliably assess the dynamical state of
the clusters from spectroscopic data alone, it would probably be
necessary to observe galaxy samples an order of magnitude larger than
those we present in this paper. In addition, X-ray observations,
particularly using Chandra and XMM, would be an efficient aid to
measure substructure and determine the dynamical state of the
clusters.  We plan to investigate the relation between weak lensing
and X-ray properties for a larger sample of clusters in a future paper
(H.\ Dahle et al.\ 2002b, in preparation).

\acknowledgements

We thank Dr.\ T.~C.\ Beers for letting us use his ROSTAT statistical
analysis program and Dr.~W.~J.\ Sutherland for providing us his
redshift measurement utility {\tt crcor}. We thank the staff of the
William Herschel Telescope, the University of Hawaii 2.2m Telescope
and the Nordic Optical Telescope for support during our observing
runs. HD gratefully acknowledges the Research Council of Norway for a
doctoral research fellowship. RJI, PBL and HD thank the Research
Council of Norway for travel support. This research has made use of
the NASA/IPAC Extragalactic Database (NED) which is operated by the
Jet Propulsion Laboratory, California Institute of Technology, under
contract with the National Aeronautics and Space Administration.

}


\begin{thebibliography}{13}
\expandafter\ifx\csname natexlab\endcsname\relax\def\natexlab#1{#1}\fi

\bibitem[Adami et al.(1998)]{Adami98} Adami, C., Mazure, A., Katgert,
P., \& Biviano, A. 1998, \aap, 336, 263
\bibitem[Allington-Smith et al.(1994)]{Allington} Allington-Smith,
J., et al.\ 1994, \pasp, 106, 983
\bibitem[Bahcall \& Tremaine(1981)]{Bahcall}Bahcall, J. N. \&
  Tremaine,  S. 1981, \apj, 244, 805 
\bibitem[Bahcall \& Cen(1993)]{Bahcall_Cen}Bahcall, N. A. \& Cen,
  R. 1993, \apjl, 407, L49
\bibitem[Bartelmann \& Scneider(2001)]{Bart_Schneid}Bartelmann, M. \&
Schneider, P. 2001, Physics Reports, 340, 291
\bibitem[Beers et al.(1990)]{Beers_et_al}Beers, T. C., Flynn, K., \&
  Gebhardt, K. 1990, \aj, 100, 32
\bibitem[B\"ohringer(2000)]{bohringer2000} B\"ohringer,  H. et
al. 2000, \apjs, 129, 435 
\bibitem[Borgani et al.(1999)]{Borgani_et_al}Borgani, S., Girardi, M.,
Carlberg, R. G., Yee, H. K. C., \& Ellingson, E. 1999, \apj, 527, 572
\bibitem[Brainerd et al.(1999)]{Brainerd_et_al}Brainerd, T. G.,
Wright, C. O., Goldberg, D. M., \& Villumsen, J. V. 1999, \apj,
524, 9
\bibitem[Briel \& Henry(1993)]{Briel&Henry}Briel, U. G. \& Henry,
  J. P. 1993, \aap, 278, 379
\bibitem[Carlberg et al.(1996)]{Carlberg96}Carlberg, R. G., Yee,
  H. K. C., Ellingson, E., Abraham, R., Gravel, P., Morris, S., \&
  Pritchet, C. J. 1996, \apj, 462, 32
\bibitem[Colless et al.(1990)]{Colless90}Colless, M., Ellis, R. S.,
  Taylor, K., \& Hook, R. N. 1990, \mnras, 244, 408
\bibitem[Dahle et al.(2002)]{Dahle00}Dahle, H., Kaiser, N., Irgens,
  R. J., Lilje, P. B., \& Maddox, S. J. 2002, \apjs, 139, 313
  (Paper I)
\bibitem[Ebeling et al.(1996)]{eb96}Ebeling, H., Voges, W.,
B{\"o}hringer, H., Edge, A.C., Huchra, J. P., \& Briel, U.G. 1996,
\mnras, 281, 799 
\bibitem[Ebeling et al.(1998)]{eb98}Ebeling, H., Edge, A. C.,
B{\"o}hringer, H., Allen, S., Crawford, C. S., Fabian, A.C., Voges,
W., \& Huchra, J. P. 1998, \mnras, 301, 881
\bibitem[Efron(1987)]{Efron}Efron, B. 1987, Journal of the American
  Statistical Association, 82, 171
\bibitem[Eke et al.(1996)]{Eke}Eke, V. R., Cole, S., \& Frenk,
  C. S. 1996, \mnras, 282, 263
\bibitem[Girardi et al.(1993)]{Girardi_et_al}Girardi, M., Biviano, A.,
Giuricin, G., Mardirossian, F., \& Mezzetti, M. 1993, \apj, 404, 38
\bibitem[Girardi \& Mezzetti(2001)]{GirardiMezzetti}Girardi, M \&
Mezzetti, M. 2001, \apj, 548, 79 
\bibitem[Gross et al.(1998)]{Gross}Gross, M. A. K., Somerville, R. S.,
  Primack, J. R., Holtzman, J., \& Klypin, A. 1998, \mnras, 301, 81
\bibitem[Hoekstra et al.(2001)]{Hoekstra}Hoekstra, H., Franx, M.,
Kuijken, K., \& van Dokkum, P. G. 2001, preprint (astro-ph/0109445) 
\bibitem[Horne(1986)]{Horne}Horne, K. 1986, \pasp, 98, 609
\bibitem[Huchra et al.(1990)]{Huchra}Huchra, J. P., Henry, J. P.,
  Postmann, M., \& Geller, M. J. 1990, \apj, 365, 66
\bibitem[Jacoby et al.(1984)]{Jacoby}Jacoby, G. H., Hunter, D. A., \&
  Christian, C. A. 1984, \apjs, 56, 257
\bibitem[Kaiser et al.(1995)]{Kaiser}Kaiser, N., Squires, G., \&
  Broadhurst, T. 1995, \apj, 449, 460
\bibitem[Kennicutt(1992)]{Kennicutt}Kennicutt, R. C. 1992, \apjs, 79,
  255
\bibitem[Lavery \& Henry(1998)]{LaveryHenry}Lavery, R. J. \& Henry,
J. P. 1998, \baas, 30, 864
\bibitem[Leir \& van den Bergh(1977)]{1977ApJS...34..381L} Leir, A.~A.~\& van den Bergh, S.\ 1977, \apjs, 34, 381 
\bibitem[Liang et al.(2000)]{Liangetal}Liang, H., L\'emonon, L.,
Valtchanov, I., Pierre, M., \& Soucail, G. 2000, A\&A, 363, 440
\bibitem[Lilje(1992)]{Lilje}Lilje, P. B. 1992, \apjl, 386, L33
\bibitem[Lilje \& Efstathiou(1988)]{Lilje_Ef}Lilje, P. B. \&
Efstathiou, G. 1988, \mnras, 231, 635
\bibitem[Limber \& Mathews(1960)]{LimberMathews}Limber, N. D. \&
  Mathews, W. G. 1960, \apj, 132, 286
\bibitem[Mellier(1999)]{mellier99}Mellier, Y. 1999, \araa, 37, 127
\bibitem[Metzger \& Ma(2000)]{metzger_ma} Metzger, M. R. \& Ma,
C.-P. 2000, \aj, 120, 2883  
\bibitem[Metzler, White, \& Loken(2001)]{2001ApJ...547..560M} Metzler,
C.~A., White, M., \& Loken, C.\ 2001, \apj, 547, 560 
\bibitem[Mosteller \& Tukey(1977)]{1977dars.book.....M} Mosteller, F.~\& Tukey, J.~W.\ 1977, Addison-Wesley Series in Behavioral Science: 
Quantitative Methods, Reading, Mass.: Addison-Wesley, 1977 
\bibitem[Navarro, Frenk, \& White(1995)]{1995MNRAS.275..720N} Navarro,
J.~F., Frenk, C.~S., \& White, S.~D.~M.\ 1995, \mnras, 275, 720
\bibitem[Olsen(2000)]{Olsen}Olsen, L. F. 2000, Ph.D. thesis,
Copenhagen University
\bibitem[Patel et al.(2000)]{Patel}Patel, S. K. et al.\ 2000, \apj,
541, 37 
\bibitem[Smail et al.(1997)]{Smail97}Smail, I., Ellis, R. S.,
Dressler, A., Couch, W. J., Oemler, A., Sharples, R. M., \& Butcher,
H. 1997, \apj, 479, 70
\bibitem[Smith(1936)]{Smith}Smith, S. 1936, \apj, 83, 23
\bibitem[Struble \& Rood(1999)]{1999ApJS..125...35S} Struble, M.~F.~\& Rood, H.~J.\ 1999, \apjs, 125, 35 
\bibitem[Tonry \& Davis(1979)]{Tonry_Davis}Tonry, J. \& Davis,
  M. 1979, \aj, 84, 1511
\bibitem[White et al.(1993)]{White_et_al}White, S. D. M.,
      Efstathiou, G., \& Frenk, C. S. 1993, \mnras, 262, 1023
\bibitem[White \& Frenk(1991)]{White_Frenk}White, S. D. M. \&
  Frenk, C. S. 1991, \apj, 379, 52
\bibitem[Yee et al.(1996)]{1996ApJS..102..289Y} Yee, H.~K.~C., Ellingson, 
E., Abraham, R.~G., Gravel, P., Carlberg, R.~G., Smecker-Hane, T.~A., 
Schade, D., \& Rigler, M.\ 1996, \apjs, 102, 289 
\bibitem[Zwicky(1933)]{Zwicky}Zwicky, F. 1933, Helv. Phys. Acta, 6,
  110

\end{thebibliography}
\end{document}